\documentclass[10pt]{article}
\usepackage{amssymb} 
\usepackage{amsmath}
\usepackage{amsthm}   
\usepackage{epsfig}
\usepackage{multirow}
\usepackage{natbib}
\usepackage{color}
 \usepackage{a4wide}
\usepackage{bm}	
\usepackage{url}	

\usepackage{booktabs}
\usepackage{placeins}
\usepackage{colortbl}
\usepackage{caption}
\usepackage{subcaption}
\usepackage{makecell}
\usepackage[dvipsnames]{xcolor}

\bibpunct{(}{)}{;}{a}{}{,}

\newtheorem{Th}{Theorem}[section]
\newtheorem{Def}[Th]{Definition}

\newtheorem{Cor}[Th]{Corollary}

\newfont{\smcal}{cmu10 scaled 1200}
\newfont{\handw}{cmmi10 scaled 1200}
\newfont{\handws}{cmmi10 scaled 800}

\newcommand{\RR}{{\mathbb R}}

\newcommand{\var}{\mbox{\rm var}}
\newcommand{\cov}{\mbox{\rm cov}}

\newcommand{\tr}{\mbox{\rm trace}}

\newcommand{\sinc}{\mbox{\rm sinc}}
\newcommand{\argmin}{\mathop{\mbox{\rm argmin}}}

\newcommand{\iid}{{\stackrel{i.i.d.}{\sim}}}
\newcommand{\R}{{\mathbb R}}%
\newcommand{\E}{{\mathbb E}}%

\newcommand{\N}{\mathbb{N}}

\newcommand{\Exp}{\mbox{\rm Exp}}
\newcommand{\Log}{\mbox{\rm Log}}
\newcommand{\Inull}{\mathcal{I}_0\big(G\big)}
\newcommand{\Diff}{\rm{Diff}^+\!(\,I\,)}

\newcommand{\so}{\mathfrak{so}}
\newcommand{\fg}{\mathfrak{g}}
\newcommand{\G}{G}

\newcommand{\Shape}{\mathcal{S}}

\newcommand{\cC}{\mathcal{C}}

\newcommand{\cI}{\mathcal{I}}

\newcommand{\Prb}{\mathbb{P}}
\newcommand{\Norm}{\mathcal{N}}

\title{Confidence Tubes for Curves on $SO(3)$ and Identification of Subject-Specific Gait Change after Kneeling}
\author{Fabian J.E. Telschow\footnote{Division of Biostatistics, University of California, San Diego}, Michael R. Pierrynowski\footnote{School
 of Rehabilitation Science, McMaster University, Canada
} and
 Stephan F. Huckemann\footnote{Felix Bernstein Institute for Mathematical Statistics in the Biosciences, Georgia Augusta University of G\"ottingen} 
 }

\begin{document}
\date{}
\maketitle

\begin{abstract}

In order to identify changes of gait patterns, e.g. due to prolonged occupational kneeling, which is believed to be major risk factor, among others, for the development of knee osteoarthritis, we develop confidence tubes for curves following a Gaussian perturbation model on $SO(3)$. These are based on an application of the Gaussian kinematic formula to a process of Hotelling statistics and we approximate them by a computable version, for which we show convergence. Simulations endorse our method, which in application to gait curves from eight volunteers undergoing kneeling tasks, identifies phases of the gait cycle that have changed due to kneeling tasks. We find that after kneeling, deviation from normal gait is stronger, in particular for older aged male volunteers.  Notably our method adjusts for different walking speeds and marker replacement at different visits.

\end{abstract}

 {\bf Keywords:} 
Functional data analysis, 
modulo group actions, 
Gaussian perturbation models, Gaussian kinematic formula, two-sample tests, Lie groups

\section{Introduction}

   There is overwhelming evidence that \emph{prolonged occupational kneeling} (POK), e.g. floor tile laying, constitutes a major risk factor for the development of knee osteoarthritis, e.g. \cite{cooper1994occupational,coggon2000occupational,rytter2009occupational}.
  Also, POK is a risk factor for the development of degenerative tears in medial menisci, e.g. \cite{rytter2009occupationalmenisci}.
    In order to identify 
    hypothesized underlying changes of gait patterns,
     kneeling workers'
     and controls' gait has been compared by \cite{gaudreault2013comparison} and prolonged kneeling has been \emph{simulated} and gait changes compared by \cite{kajaks2015effect,tennant2018effects}. Also, dependence of kneeling effects due to footwear has been investigated by \cite{tennant2015effect} and
    kneeling effects  have been studied on cadavers with total knee arthroplasty \cite{wilkens2007biomechanical}.
    
    In order to assess the specifics of changes of gait patterns, the three dimensional rotational path in $SO(3)$ of the relative motion of the tibia (larger lower leg bone) w.r.t. the femur (upper leg bone) is usually represented by the three Euler angles flexion/extension, adduction/abduction
    and internal/external rotation. Doing so, \cite{gaudreault2013comparison, kajaks2015effect,tennant2018effects} have found, among others, for each angle, loci of significant gait changes, without, however, addressing the issue of multiple testing, correlation of the sequential data and the effect of marker replacement.
    
    In our approach, we address all of these issues, and in consequence, are able to test for subject-specific changes of gait pattern. In application, we do this for pre- and post-kneeling, the method, however, is applicable for any change of condition (e.g. onset of otheoathritis) over a 
    period of time, due to correcting 
    for marker replacement. To this end, we recall a Gaussian perturbation model from \cite{TelschowHuckemannPierrynowski2016} for curves on Lie groups and show that a Hotelling statistic for the corresponding process follows asymptotically (for vanishing variance) a Hotelling statistic that can be described by a \emph{Gaussian kinematic formula} (GFK) from \cite{TaylorTakemuraAdler2005,Taylor2006}. In application to gait analysis, our method, relying on curves on the rotational group, takes advantage of simultaneously involving all three Euler angles in a canonical way. Moreover, as our test statics use maxima of stochastic processes, we resolve the multiple testing issue by providing for simultaneous confidence tubes over entire gait cycles. Further, sequential correlation is naturally modeled within the GFK approach by simulating quantiles from the empirical process.
    Indeed, simulations mimicking and going beyond the use case of low variance and high smoothness typical in gait analysis show that our method is well applicable. Then, for an experiment conducted in the School of Rehabilitation Science at McMaster University (Canada), for six out of eight  healthy volunteers we identify individual changes of gait patterns after kneeling tasks.
    We find that after kneeling, deviation from normal gait is stronger, in particular for older aged male volunteers. 
    
    Remarkably, our method is also robust under specialist marker replacement, at a subsequent patient's visit, say. It is well known that Euler angle curves may considerable change after marker replacement and simple approaches subtracting average angles over gait cycles (cf. \cite{kadaba1989repeatability}) have remained questionable, e.g. \citet{DelvalAlleronETAL08, mcginley2009reliability,noehren2010improving,roislien2012evaluating}, also for other approaches, leading to the longstanding open problem of \emph{gait reproducibility}, see \citet{duhamel2004statistical}.
    
    In a recent publication (\cite{TelschowHuckemannPierrynowski2016}), we have developed a method to successfully correct for marker replacement by estimating a Lie group isometry, bringing two samples (each sample is a repeated measurement of the same person's gait with fixed marker placement) into optimal position to one another. Since volunteers will have different comfortable walking speeds at different visits, we have also corrected for a sample-specific time warping effect. This method is part of the tool chain developed in the present contribution which is available under
     \url{www.stochastik.math.uni-goettingen.de/KneeMotionAnalytics}
    as an R-package. In particular, it contains all data and code used in this paper.

    \section{Testing Gaussian Perturbation Models on Lie Groups Modulo Sample-Specific Spatio-Temporal Action}\label{scn:GPMs} 
    The following is taken, 
    from \cite{TelschowHuckemannPierrynowski2016}. It has been formulated for $p=1$ and generalizes at once to arbitrary $p\in \N$. Let $G$ be a connected Lie group with Lie algebra $\fg$ embedded in a suitable Euclidean space $\RR^m$ and Lie exponential $\Exp : \fg \to G$. With the unit interval $I=[0,1]$ we have the family $\Gamma =  \cC^p([0,1],G)$ of $p \in \N$ times continuously differentiable curves on $G$. 
    We assume in particular that $G$ admits a bi-invariant Riemannian metric, a sufficient condition for which is that $G$ is compact.
    
    \begin{Def}
          We say that a random curve $\gamma\in \Gamma$ follows a \emph{Gaussian perturbation} (GP) around a \emph{center curve} $\gamma_0\in \Gamma$ if there is a $\fg$-valued zero-mean Gaussian process $A_t$ with a.s. $\cC^p$ paths, such that
      \begin{equation}\label{eqdef:rGP}
	    \gamma(t)  = \gamma_0(t)\Exp\big( A_t\big)\mbox{ for all }t\in I\,.
      \end{equation}
       The Gaussian process $A_t$ will be called the \emph{generating process}. 
     \end{Def}
     
     This model, which is based on right multiplication with the exponential of the generating process is equivalent to one based on left multiplication and asymptotically (as the variance goes to zero) equivalent to one based on two-sided multiplication, cf. \cite{TelschowHuckemannPierrynowski2016}. 
     Moreover, this model is invariant under the \emph{spatial} action of the isometry group $\cI(G)$ on $G$, where its connectivity component $\Inull$ of the identity element can be viewed as the analog of the orientation preserving Euclidean motions of a Euclidean space. 
     Indeed, for $G$ compact, semisimple with trivial center, we have $G\times G = \Inull$, cf. \cite{TelschowHuckemannPierrynowski2016}, which is the case for $G= SO(3)$.
     
     Also, Model (\ref{eqdef:rGP}) is invariant under the temporal action 
     $$\phi \in \Diff = \big\{ \phi \in \mathcal{C}^\infty( I,I): \phi'(t) > 0~\text{ for all } t\in I\big\}$$
     of strictly monotone time warpings. We set $\Shape = \Inull \times \Diff$ and write $(\psi,\phi): \gamma \mapsto \psi\circ\gamma\circ \phi$ for the corresponding action on $\Gamma$, $\psi\in \Inull$ and $\phi \in \Diff$.
     
     If $G$ admits a bi-invariant Riemannian metric (it does, if it is compact, say), on $\Gamma$ we introduced the \emph{intrinsic length loss} 
     \begin{equation*}
      \delta(\gamma,\eta) = \tfrac{1}{2}\big(\delta_{1}(\gamma,\eta) + \delta_{2}(\gamma,\eta)\big)\,.
     \end{equation*}
    where 
    \begin{align*}
	\delta_{1}(\gamma,\eta) &= {\rm length}(\gamma\eta^{-1})~ ~ \text{ and }  ~ ~ \delta_{2}(\gamma,\eta) = {\rm length}(\gamma^{-1}\eta)\,,
    \end{align*} 
    for $\gamma,\eta \in \Gamma$. Here, the length is taken with respect to the bi-invariant metric on $\G$,
       \begin{align*} 
       {\rm length}(\gamma) = \int_0^1 \|\dot \gamma(t)\|\,dt 
       \end{align*}
    The loss $\delta$ is invariant under the spatio-temporal action. There are other loss functions, canonical on Euclidean space modulo time warping, that can be extended to manifolds, cf. \cite{SrivastavaWuKurtekKlassenMarron2011,SuKurtekKlassenSrivastava2014}. 
     
    For independent i.i.d. samples $\chi_1=\{\gamma_1,...,\gamma_N\}$ and $\chi_2=\{\eta_1,...,\eta_M\}$, $N,M\in \N$, of GP models $\gamma$ and $\eta$ with center curves $\gamma_0$ and $\eta_0$, respectively, we have developed in \cite{TelschowHuckemannPierrynowski2016} rank permutation tests for
    \begin{equation}\label{eq:NullHypoRGP}
    H_0:~~\exists (\psi,\phi)\in \Shape:~ \gamma \sim \psi\circ\eta\circ\phi~~~ ~vs.~~~ ~H_1:~~\forall (\psi,\phi) \in \Shape:~ \gamma\not\sim \psi\circ\eta\circ \phi
    \end{equation}
    at a given significance level $\alpha\in(0,1)$. Notably, in contrast to classical shape analysis correcting for group action on individual measurements, we correct for a common \emph{sample-specific group action} and to this end, in application in Section \ref{scn:Application}, we apply \citet[Test 2.11]{TelschowHuckemannPierrynowski2016}.

\section{Confidence Tubes on $G$}\label{scn:Tubes}

    Since $G$ is connected by hypothesis, the inverse exponential is well defined on the complement in $G$ of the cut locus of the unit element. 
    Let $\Log : G \to \fg$ denote a measurable extension. Further, since $\fg$ is a linear space, let $\iota:\RR^m \to \fg$ be a suitable isomorphism and set
    $\mathfrak{L} = \iota^{-1}\circ \Log : G \to \RR^m$.

    \begin{Def}\label{def:ResidualProcess}
    Let $\gamma_1,\ldots,\gamma_N$ be a sample of a random curve $\gamma \in \Gamma$ following a GP model around a center curve $\gamma_0$ and let $\hat \gamma_{N}$ be an estimator for $\gamma_0$. Then
    \begin{align}
	x_t^N	&= \mathfrak{L}\Big(\,  \hat \gamma_N^{-1}(t) \gamma_{0}(t)\, \Big)\,, 
	\\
	x^{N,n}_t 	&= \mathfrak{L}\Big(\, \hat \gamma_{N}^{-1}(t) \gamma_{n}(t)\, \Big)
	\end{align}
	are called \emph{intrinsic population} and  \emph{sample} \emph{residuals}, respectively. 
    \end{Def}

    This gives rise to the following one-dimensional processes,
    \begin{equation}\label{eq:stochasticprocessforknees}
	\hat H^{x,N}_t = N(x^N_t)^T \Big(\hat S^{x,N}_t \Big)^{-1}\! x^N_t\,, ~\text{ where }~ ~ ~\hat S^{x,N}_t = \frac{1}{N-1}
	\sum_{n=1}^N x^{N,n}_t(x^{N,n}_t)^T\,,	
    \end{equation}
    where we assume that $\hat S^{x,N}_t$ is non-singular for all $t\in[0,1]$. Further, for $0\leq \alpha \leq 1$  we define the quantile
    \begin{equation*}
	\hat h_{\gamma,N,\alpha} = \inf \left\{ h\in\R_{\geq0}\,\big\vert~\Prb\left\{ \sup_{t\in [0,1]} \hat H^{x,N}_t \leq h \right\} \geq 1-\alpha\right\}\,.
    \end{equation*}
    From this we obtain at once simultaneous $(1-\alpha)$-confidence tubes for $\gamma_0$, setting
	\begin{equation*}
	      \mathcal{V}_{\alpha}\big(\gamma_1,\ldots,\gamma_N; t\big) = \left\{a\in \fg\,\big\vert~ Na^T\Big(\hat S^{x,N}_t \Big)^{-1}\! a\leq \hat h_{\gamma,N,\alpha} \right\}\,.
	\end{equation*}
    \begin{Th}\label{theorem:SO3scs}
	Let $\gamma_1,\ldots,\gamma_N$ be a sample of a random curve $\gamma \in \Gamma$ following a GP model around a center curve $\gamma_0$. Let $\hat \gamma_{N}$ be an estimator for $\gamma_0$ and assume $\hat S^{x,N}_t$ is non-singular for all $t\in[0,1]$. 
	Then
	\begin{equation*}
	      \Prb\Big\{\, \gamma_0(t) \in \hat \gamma_{N}(t)~\Exp\Big(\iota \circ\mathcal{V}_{\alpha}\big(\gamma_1,\ldots,\gamma_N; t\big) \Big) \text{ for all }t\in [0,1]\, \Big\} \geq \alpha
	\end{equation*}
	and hence this set forms a simultaneous $(1-\alpha)$-confidence tube for $\gamma_0$.
    \end{Th}
 
 The process $\hat H^{x,N}_t$ from (\ref{eq:stochasticprocessforknees}) serves as an approximation of the \emph{genuine Hotelling process} determined by (\ref{eqdef:rGP}): 
	\begin{equation}\label{genuine-H:eq}
  H^{a,N}_t = N\, 
  (\bar a^N_t)^T (S^{a,N}_t)^{-1} \bar a^N_t\,
 \end{equation}
 where $\gamma_n(t) = \gamma_0\,\Exp(A_t^n)$, $\E[A_t^n] = 0$ for $1\leq n\leq N$ and $t\in I$, as well as, 
	\begin{equation}\label{genuine-res:eq}
	\iota^{-1}\circ A_t^n = a_t^n\,,\quad
 \overline{a}^N_t = \frac{1}{N}\sum_{n=1}^N a^n_t\quad\mbox{ and }\quad S_t^{a,N} = \frac{1}{N-1}\sum_{n=1}^M(a_t^n - \overline{a}^N_t)(a_t^n - \overline{a}_t^N)^T\,.
 \end{equation}
 Among others, the following section makes this approximation explicit for the special case of $G=SO(3)$.
 
 \section{GP Models and Approximating Confidence Tubes on $SO(3)$}
 
 For $\G$, the compact and connected Lie group of three-dimensional rotations $G=SO(3)$ we detail the above approximation. To this end, we first recall the structure of $SO(3)$, extrinsic pointwise means as estimators $\hat\gamma_N$ and fundamental properties of corresponding GP models.  
 
 \subsection{GP Models on $SO(3)$}
 
 The Lie group $G= SO(3)$
 comes with the Lie algebra $\fg=\so(3)=\{A\in \R^{3\times 3}: A^T=-A\}$ of $3\times 3$ skew symmetric matrices. This Lie algebra is a three-dimensional linear subspace of all $3\times 3$ matrices and thus carries the  natural structure of $\R^3$ conveyed by the isomorphism $\iota: \mathbb R^3 \rightarrow \mathfrak{so}(3)$ given by
\begin{equation*} 
 \iota(a) = \begin{pmatrix}
  0 & -a_3 & a_2 \\
  a_3 & 0 & -a_1 \\
  -a_2 & a_1 & 0
 \end{pmatrix},\quad \mbox{ for } a = (a_1,a_2,a_3)^T \in \R^3\,.
 \end{equation*}
 This isomorphism exhibits at once the following relation
 \begin{equation}\label{iota-rotation:eq}
  Q \iota(a) Q^T= \iota(Qa)\mbox{ for all }a\in \R^3\mbox{ and }Q\in G\,.
 \end{equation}
We use the scalar product $\langle A,B \rangle:=\tr\big(AB^T\big)/2 = a^Tb$ for $\iota^{-1}(A)=a$ and $\iota^{-1}(B)=b$, which induces the rescaled Frobenius norm $\Vert A\Vert_F = \sqrt{\tr(AA^T)/2} = \|a\|$,  on all $\RR^{3\times 3}$. 
On $G\subset \RR^{3\times 3}$ it induces the \emph{extrinsic metric}, cf. \cite{BP03}. Moreover, we denote with $I_{3\times 3}$ the unit matrix. As usual, $A\mapsto \Exp(A)$ denotes the matrix exponential which is identical to the Lie exponential and gives a surjection $\fg \to G$. Due to skew symmetry, the following \emph{Rodriguez formula} holds
\begin{equation}\label{eq:rodriguez}
 \Exp(A)  = \sum_{j=0}^\infty \frac{A^j}{j!}
	  = I_{3\times 3} + \frac{\sin(\Vert A \Vert_F)}{\Vert A \Vert_F}\,A + \frac{1-\cos(\Vert A \Vert_F)}{\Vert A \Vert_F^2}\,A^2\,.
\end{equation}
This yields that the Lie exponential is bijective on $\mathcal{B}_\pi(0) = \{A\in \fg: \|A\|_F < \pi\}$.  For a detailed discussion, see \cite[p. 121]{ChirikjianKyatkin2000}.

As in \cite{TelschowHuckemannPierrynowski2016}, introduce \emph{pointwise extrinsic mean} (PEM) curves $\hat \gamma_N(t)$ of a sample $\gamma_1,\ldots,\gamma_N\iid\gamma\in \Gamma$, which are defined for each $t\in[0,1]$ by
\begin{equation}
  \hat \gamma_N(t) \in \hat E_N(t) = \argmin_{\mu\in \G}\frac{1}{N}\sum_{n=1}^N  \Vert \mu - \gamma_n(t) \Vert_F^2\,.
\end{equation}
They fulfill the following uniqueness and convergence properties for GP models as proven in \citet{TelschowHuckemannPierrynowski2016}.
\begin{Th}\label{theorem:PESMuniformconvergence}
    Let $\gamma_1,\ldots,\gamma_N$ be a sample of a random curve $\gamma \in \Gamma$ following a GP model around a center curve $\gamma_0$ and let $t\mapsto \hat \gamma_N(t)$ be a measurable selection of $\hat E_N(t)$ for each time point $t\in [0,1]$. If the generating Gaussian process $A_t$  satisfies
 \begin{equation}\label{eq:uniformAssumption}
    \E\left[ \max_{t\in [0,1]} \Vert \partial_t A_t \Vert_F \right] <\infty\,,
 \end{equation}
then the following hold.
 \begin{enumerate}
  \item[(i)] There is $\Omega'\subset \Omega$ measurable with $\Prb(\Omega')=1$ such that for every $\omega\in \Omega'$ there is $N_\omega\in \mathbb N$ such that for all $N\geq N_\omega$, every $\hat E_N(t)$ has a unique element $\hat \gamma_N(t)$, for all $t\in [0,1]$, and $\hat \gamma_N \in \Gamma$;
  \item[(ii)] $\max_{t\in [0,1]} \big\Vert \hat\gamma_N(t) - \gamma_0(t)\big\Vert_F \rightarrow 0$ 
  for $N\rightarrow\infty$ \mbox{almost surely.}
 \end{enumerate}
\end{Th}
\begin{Cor}\label{corollary:PESMoftenContinuous}
With the notations and assumptions of Theorem \ref{theorem:PESMuniformconvergence} we have
 \begin{equation*}
  \lim_{N\rightarrow\infty} \Prb\big\{ t \mapsto \hat\gamma_N(t)\in \Gamma \big\} = 1\,.
 \end{equation*}
\end{Cor}

\subsection{Approximating Confidence Tubes on $SO(3)$}

  As the main result of this section we first show that in case of concentrated errors (as is typical in biomechanics, e.g. \citet{RancourtRivestAsselin2000}), the residual processes $x^N_t$ and $X^{N,n}_t$ from Definition \ref{def:ResidualProcess} are approximatively the residuals of the generating Gaussian process (\ref{genuine-res:eq}) of the GP model. Then, we  use this approximation to define an estimator for $\hat h_{\gamma,N,\alpha}$ based on the Gaussian kinematic formula for Hotelling processes (see \citet{TaylorWorsley2008}), which will be shown in Section \ref{scn:Simulations}, using simulations, to perform very reliably even if the sample sizes are small as it is usually the case in biomechanical gait analysis.

    \begin{Th}[Approximations for Concentrated Errors]\label{theorem:residuals}
	  Let $N\in \N$ be fixed and $\gamma_1,\ldots,\gamma_N$ be a sample of a random curve $\gamma \in \Gamma$ following a GP model around a center curve $\gamma_0$. Additionally, assume that the generating Gaussian process $A_t = \iota\circ a_t$ satisfies $\E\left[ \max_{t\in [0,1]} \Vert \partial_t A_t \Vert_F \right] <\infty$ and $\max_{t\in [0,1]} \Vert A_t \Vert_F = \mathcal{O}_p(\sigma)$ with $0<\sigma\rightarrow0$. Let $\hat\gamma_{N}(t)$ be a measurable selection of sample PEM curves. Then, for $x^N_t$ and $x^{N,n}_t$ from Definition \ref{def:ResidualProcess},
	  \begin{align}\label{eq:residualsClaim1}
		  x^N_t &= \bar{a}^N_t  + \mathcal{O}_p\left( \sigma^2 \right)\\\label{eq:residualsClaim2}
		  x^{N,n}_t &= a^{n}_t - \bar{a}^N_t + \mathcal{O}_p\left(\sigma^2 \right),
	  \end{align}
	  where $\mathcal{O}_p\!\left(\sigma^2 \right)$ is uniform over $t\in[0,1]$.
    \end{Th}

    \begin{Cor}[Asymptotically genuine Hotelling process]\label{corollary:Tapproximation}
      With the assumptions and notations of Theorem \ref{theorem:residuals} we obtain with $\mathcal{O}_p\!\left(\sigma \right)$ uniformly over $t\in [0,1]$,
      \begin{equation*}
	H^{a,N}_t = \hat H^{x,N}_t + \mathcal{O}_p(\sigma)\,,
      \end{equation*}
      if additionally $\cov\big[a_t\big]= \sigma^2\Sigma_t$ with fixed and non-singular $\Sigma_t$ for all $t\in [0,1]$.
    \end{Cor}

     The following theorem gives the equivariance property of the simultaneous confidence tubes with respect to the group action on $\Gamma$ of the group $\Inull\times\Diff$, which is $(G\times G)\times\Diff$ by Section \ref{scn:GPMs}.
    \begin{Th}\label{theorem:EquivarianceConfidenceSets}
	Let $\gamma_1,\ldots,\gamma_N$ be a sample of a random curve $\gamma \in \Gamma$ following a GP model around a center curve $\gamma_0$ with PEM curve $\hat \gamma_{N}$. Moreover, let $(\psi,\phi) \in (G\times G)\times\Diff$ be arbitrary and define the sample $\eta_n=\psi\circ\gamma_n\circ \phi$, $n\in\{1,\ldots, N\}$ of the GP $\psi\circ\gamma\circ\phi$ with center curve $\eta_0=\psi\circ\gamma_0\circ\phi$ and PEM curve $\hat \eta_{N}$.
	Then, for every $0\leq \alpha\leq 1$, the simultaneous confidence tubes for $\psi\circ\gamma_0\circ\phi$ computed from $\eta_1,\ldots,\eta_N$ satisfy
	\begin{equation*}
	    \hat\eta_{N}(t)~ \Exp\Big( \iota\circ\mathcal{V}_{\alpha}\big(\eta_1,\ldots,\eta_N; t\big) \Big) =  (\psi\circ\hat\gamma_{N}\circ\phi)(t) ~\Exp\Big( \iota\circ Q_{\psi}\mathcal{V}_{\alpha}\big(\gamma_1,\ldots,\gamma_N; \phi(t)\big) \Big)\,,
	\end{equation*}
	i.e., they can be derived from the simultaneous confidence tubes for $\gamma_0$ using $\gamma_1,...,\gamma_N$ and $(\psi,\phi) \in (G\times G)\times\Diff$ only.
    \end{Th}

    \paragraph{The Gaussian kinematic formula (GKF).}
      Corollary \ref{corollary:Tapproximation} states that for concentrated errors the statistic $H^a_t$, which is the Hotelling $T^2$ statistic of a generating Gaussian process, approximates the statistic $\hat H^{x,N}_t$. Thus, in order to estimate the quantiles $\hat h_{\gamma,N,\alpha}$ for the process $\hat H^{x,N}_t$, derived from a GP model $\gamma$, we use the \emph{expected Euler characteristic heuristic} (see \citet{TaylorTakemuraAdler2005}) and assume that
      \begin{equation}\label{eq:eulerHeuri}
	  \Prb\Big( \max_{t\in [0,1]} \hat H^{x,N}_t > h \Big) \approx  \E\!\left[ \mathfrak{x}\Big(\big\{ t\in [0,1]\,\vert~ \hat H^{x,N}_t \geq h\big\} \Big) \right] \approx  \E\!\left[ \mathfrak{x}\Big(\left\{ t\in [0,1]\,\vert~H^{a,N}_t \geq h\right\} \Big) \right]\,,
      \end{equation}
      where $\mathfrak{x}(\mathcal{U})$ denotes the \emph{Euler characteristic} (EC) of $\mathcal{U}\subset [0,1]$. Although we cannot rigorously justify this approximation, our simulations in Section \ref{scn:Simulations} show that this procedure works very well.

      Under some additional technical assumptions on the generating Gaussian process $A_t = \iota\circ a_t$ given in \citet{Taylor2006}, it is shown in \citet{TaylorWorsley2008} that the expected EC of the excursion set $\left\{t\in [0,1]\,\vert~H^{a,N}_t\geq h\right\}$ can be computed explicitly by the formula
      \begin{equation}\label{equation:GKF}
	  \E\Big[\mathfrak{x}\!\left\{t\in [0,1]\,\vert~H^{a,N}_t\geq h\right\}\Big] = \mathcal{L}_0\big([0,1]\big) \rho^{H}_0(h) + \mathcal{L}_1\big([0,1]\big) \rho^{H}_1(h)
      \end{equation}
      with the so called Lipschitz-Killing curvatures
      \begin{equation*}
	  \mathcal{L}_0\big([0,1]\big) = 1,\quad \mathcal{L}_1\big([0,1]\big) = \int_0^1 \sqrt{ \var\Big[ \tfrac{d a}{dt}(t)\Big]} dt\,.
      \end{equation*}
      
      The so called \emph{Euler characteristic densities} $\rho^H_j$ for $j\in\{1,2\}$ appearing in the GKF \eqref{equation:GKF} can be computed from the EC densities of a $T$-process with $N-1$ degrees of freedom via Roy's union intersection principle (cf{.} \citet[Sec. 3.1.]{TaylorWorsley2008}) using the formula
      \begin{equation*}
	  \rho^{H}_j(h) = \sum_{d=1}^3 \mu_d(S^2) \rho_{j+d}^{T}\big(\sqrt{h}\big)\,,~ ~ ~ ~ j=0,1\,.
      \end{equation*}
      Here $\mu_d(S^2)$ denotes the $d$-dimensional intrinsic volume of the two-sphere $S^2$ given by
      \begin{equation*}
	  \mu_0\big(S^2\big) = 2,\quad \mu_1\big(S^2\big)=0=\mu_3\big(S^2\big),\quad \mu_2\big(S^2\big) = 4\pi\,,
      \end{equation*}
      in \citet[p. 23]{TaylorWorsley2008}. In relation to the Stochastic Geometry literature, $\mu_0$ gives twice the number of connected components and $\mu_2$ gives the surface area of $S^2$ (e.g., \citet[p. 100]{MeckeStoyan2000}). Moreover, the EC densities of a $T$-process with $(N-1)$ degrees of freedom have the explicit representations
      \begin{align*}
	  \rho^T_0(t) &= \int_t^\infty \frac{\Gamma\left( \tfrac{N}{2} \right)}{\sqrt{N-1\pi}\Gamma\left( \tfrac{N-1}{2} \right)}\left( 1 + \tfrac{u^2}{N-1} \right)^{-N/2} du\\
	  \rho^T_1(t) &= \left(2\pi\right)^{-1} \left( 1 + \tfrac{t^2}{N-1} \right)^{1-N/2}\\
	  \rho^T_2(t) &= \left(2\pi\right)^{-3/2}\frac{\Gamma\left( \tfrac{N}{2} \right)}{\sqrt{\tfrac{N-1}{2}}\Gamma\left( \tfrac{N-1}{2} \right)}t\left( 1 + \tfrac{t^2}{N-1} \right)^{1-N/2}\\
	  \rho^T_3(t) &= \left(2\pi\right)^{-2} \left( \tfrac{N-2}{N-1}t^2 - 1 \right) \left( 1 + \tfrac{t^2}{N-1} \right)^{1-N/2}\,,
      \end{align*}
      given in \citet[p. 915]{TaylorWorsley2007}.

  \paragraph{Estimation of the quantile $\hat h_{\gamma,N,\alpha}$.}
  Using the GKF for Hotelling $T^2$-processes together with the EC heuristic \eqref{eq:eulerHeuri} yields
  \begin{equation*}
      \Prb\Big( \max_{t\in [0,1]} \hat H^{x,N}_t > h \Big)
	    \approx 2\rho^T_0\big(\sqrt{h}\big) - 4\pi\rho^T_2\big(\sqrt{h}\big) - \mathcal{L}_1\big([0,1]\big)\left( 2\rho^T_1\big(\sqrt{h}\big) + 4\pi\rho^T_3\big(\sqrt{h}\big) \right)\,,
  \end{equation*}
  which can be used if $\mathcal{L}_1\big([0,1]\big)$ is known, to estimate the value $\hat h_{\alpha,N,\alpha}$ for low probabilities $\alpha$ by solving
  \begin{equation}\label{equation:GKFapproximation}
  2\rho^T_0(\sqrt{h}) - 4\pi\rho^T_2(\sqrt{h}) - \mathcal{L}_1\big([0,1]\big)\left( 2\rho^T_1(\sqrt{h}) + 4\pi\rho^T_3(\sqrt{h}) \right) = 1-\alpha\,.
  \end{equation}
  Thus, it remains to estimate the Lipschitz-Killing curvature $\mathcal{L}_1\big([0,1]\big)$. This has been achieved for Gaussian processes in $\R^D$, $D\in \N$, in \citet[Sect. 4]{TaylorWorsley2007} and \citet{TaylorWorsley2008}, where they also proved that their estimator is consistent.

  By Theorem \ref{theorem:residuals} the intrinsic residuals of a sample from a GP model $\gamma$ are, in case of concentrated errors, close to the residuals of the generating Gaussian process $A_t = \iota\circ a_t$. Since the estimator of \citet[Equation (18)]{TaylorWorsley2008} is based only on the Gaussian residuals, we adapt their estimator by replacing their residuals by the intrinsic residuals given in Theorem \ref{theorem:residuals} to obtain an estimator of the Lipschitz-Killing curvature $\mathcal{L}_1\big([0,1]\big)$.

  For convenience we restate the resulting estimator. Let $\gamma_1,\ldots,\gamma_N$ be a sample of a GP model $\gamma$ and assume the curves are observed at times $0=t_1< t_2<...<t_K=1$. Then we define the matrix
  \begin{equation*}
  R_{t_k} = \left( X^1_{t_k}, \ldots, X^N_{t_k} \right)^T\in \mathbb{R}^{N \times 3}\,.
  \end{equation*}
  Further, denote by $R^d_{t_k}$ the $d$-th column of $R_{t_k}$ and define the normalized residuals as
  \begin{equation*}
  \hat R^d_{t_k} = \frac{R^d_{t_k}}{\Vert R^d_{t_k} \Vert}
  \end{equation*}
  for $d\in\{1,2,3\}$ and $k\in\{1,\ldots,K\}$. The estimator of the Lipschitz-Killing curvature $\mathcal{L}_1\big([0,1]\big)$ is then given by
  \begin{equation}\label{eq:LipEstimation}
    \hat{\mathcal{L}}_1(I) = \frac{1}{3}\sum_{k=1}^{K-1} \sum_{d=1}^{3} \big\Vert \hat R^d_{t_{k+1}} - \hat R^d_{t_k} \big\Vert\,.
  \end{equation}

\section{Simulations of Covering Rates}\label{scn:Simulations}

  Since the estimation of the quantile $\hat h_{\gamma,N,\alpha}$ relies on an approximation for concentrated error processes given in Theorem \ref{theorem:residuals}, we study the actual covering rate of this method using simulations.

  \paragraph{GP models used for simulation.}
    Without loss of generality we may assume that our center curves satisfy $\gamma_0(t)=I_{3 \times 3}$ for all $t\in[0,1]$. Otherwise, multiply the sample with $\gamma_0(t)^{-1}$.
    

    In our simulations studying the covering rates of the simultaneous confidence sets given in Theorem \ref{theorem:SO3scs}, we use the error processes
    \begin{align}
    \varepsilon^{1,l}_{t} &= f_l(t)\Big( b_1\sin\left( \tfrac{\pi}{2}t \right) + b_2\cos\left( \tfrac{\pi}{2}t \right) \Big)\nonumber\\ 
    \varepsilon^{2,l}_{t} &= f_l(t)\left( \frac{ \sum_{i=1}^{10} b_i e^{-\frac{\left(x-\frac{i-1}{9}\right)^2}{0.2}} }{\sqrt{ \sum_{i=1}^{10} e^{-2\frac{\left(x-\frac{i-1}{9}\right)^2}{0.2}} }} \right)\label{eq:SimErrorProcess}\\
    \varepsilon^{3,l}_{t} &= f_l(t)\left( b_0 e^{-5t} + \sqrt{10} \int_{0}^t e^{5(s-t)} dW_t \right)\nonumber
    \end{align}
    with i.i.d. $b_i\sim\Norm(0,1)$ for $i\in\{0,...,10\}$, $\{W_t\}_{t\in I}$ a Wiener process, and for $l\in\{1,2,3\}$ we set
    \begin{equation*}
    f_1(t) = 1\,,~ ~ ~ ~ f_2(t) = 4\,,~ ~ ~ ~ f_3(t) = \sin(4\pi t) + 1.5\,.
    \end{equation*}
    Note that the processes satisfy $\var\big[\varepsilon^{\nu,l}_{t}\big]=f_l(t)^2$ for all $t\in [0,1]$, $l\in\{1,2,3\}$ and $\nu\in\{1,2,3\}$. Moreover, the sample paths of the processes $\varepsilon^{1,l}$ and $\varepsilon^{2,l}$ have $\mathcal{C}^\infty$ sample paths, whereas the sample paths of $ \varepsilon^{3,l}$, which is a Ornstein-Uhlenbeck process (e.g., \citet[p.43]{Iacus2009}), are only continuous, implying that the GKF is not applicable for this process.
    
    From these error processes the generating Gaussian process $A_t$ of the GP model is constructed by the following formula
    \begin{equation}\label{eq:simerrorRGPprocess}
    A^{i,l,j,\sigma}_t = M_j \big(\sigma\varepsilon^{i,l}_{1,t},\sigma\varepsilon^{i,l}_{2,t},\sigma\varepsilon^{i,l}_{3,t}\big)^T\,,
    \end{equation}
    for $i\in\{1,2,3\}$, $j \in \{ 1,2 \}$, $l \in \{ 1,2,3 \}$ and $\sigma\in\R_{>0}$. Here we denote with $\varepsilon^{i,l}_{s,t}$ for $s=1,2,3$ independent realizations of $\big\{\varepsilon^{i,l}_{t} \big\}_{t\in I}$. The matrices
    \begin{equation*}
	M_1 = \begin{pmatrix}
		    1 		& 0 		& 0 \\
		    0	 	& 1	 	& 0 \\
		    0 		& 0 		& 1
	      \end{pmatrix}\,, ~ ~ ~
	M_2 = \begin{pmatrix}
		    1 		& 0 		& 0 \\
		    \frac{1}{2} 	& \frac{1}{2} 	& 0 \\
		    \frac{1}{\sqrt{3}} 		& \frac{1}{\sqrt{3}} 		& \frac{1}{\sqrt{3}}
	      \end{pmatrix}\,.
    \end{equation*}
    are introduced to include correlations among the coordinates. Moreover, (\ref{eq:simerrorRGPprocess}) introduces different variances in the coordinates, since for $j=2$ the second component has half the variance of the other two components.

  \paragraph{Design of simulation of simultaneous confidence tubes (SCTs) for center curves of GP models.}
    First, $N\in\{10,15,30\}$ realizations of the process $\{A^{i,l,j,\sigma}_t\}$ on the equidistant time grid $\mathcal{T}$ with $\Delta t= 0.01$ of $[0,1]$ for $i\in\{1,2,3\}$, $j \in \{ 1,2 \}$, $l \in \{ 1,2,3 \}$ and $\sigma\in\{0.05, 0.1, 0.6\}$ are simulated. We only report small sample sizes here, since the asymptotic behavior has been studied intensely in \citet{TelschowSchwartzman2019} and small simulation studies for higher sample sizes did not reveal departures from correct covering rates. 
    
    Then $(1-\alpha)$-SCT are constructed using Theorem \ref{theorem:SO3scs}. Here the quantile $\hat h_{\gamma,N,\alpha}$ is estimated by equation \eqref{equation:GKFapproximation} using the estimator \eqref{eq:LipEstimation} for the Lipschitz killing curvature. Afterwards it is checked whether $\gamma_0\equiv I_{3\times3}$ is contained in the SCT for all $t\in\mathcal{T}$. This procedure is repeated $M=5000$ times. The true covering rate is approximated by the relative frequency of the numbers of simulations, in which  the constructed SCT contained the true center curve $\gamma_0$ for all $t\in \mathcal{T}$.

  \paragraph{Results of simulation of SCT for center curves of GP models.}  \FloatBarrier
    The results are reported in Table \ref{tab:results} and they convey a positive message:
    For a variance $\sigma = 0.05$, which is that of the data of the application in Section \ref{scn:Application}, the simulated covering rate is very close to $(1-\alpha)$. 
    Only in the case of the Ornstein-Uhlenbeck error process we have slightly too high covering rates. For higher variance ($\sigma = 0.6$) we underestimate the covering rate. This is expected, since the proposed estimator is designed for concentrated data and the map $v\mapsto\Log\big(\Exp(v)\big)$ is only the identity on $\Vert v \Vert < \pi$ and we have the inequality
    \begin{equation}
    \Big\Vert \Log\Big(\Exp\big(\iota(v)\big)\Big) \Big\Vert_F \leq \Vert v \Vert\,.
    \end{equation}
    This implies that our estimated covariance matrix has smaller eigenvalues then the covariance matrix of the sample and hence our confidence sets will become smaller. This effect is more visible if the sample size is large, since more curves cross the cut locus.

    \begin{table}[ht]
\begin{tabular}{|llll|lll|}
  \hline
  \multicolumn{1}{|c}{$\mathbf{N}$} & \multicolumn{1}{c}{$\bm{\sigma}$} & \multicolumn{1}{c}{\textbf{E.P.}} & \multicolumn{1}{c|}{$\bm{1-\alpha}$} & \multicolumn{1}{c}{$\bm{i=1}$} & \multicolumn{1}{c}{$\bm{i=2}$}  & \multicolumn{1}{c|}{$\bm{i=3}$} \\ 
  \hline
   10 & 0.05 & $A^{i,1,1,\sigma}$ & 85/90/95 & 86.1/91.0/95.0 & 85.3/90.1/95.6  & 90.4/93.9/96.6 \\ 
   15 & 0.05 & $A^{i,1,1,\sigma}$ & 85/90/95 & 85.0/90.1/95.4 & 85.7/90.7/94.9  & 89.4/93.0/96.6  \\ 
   30 & 0.05 & $A^{i,1,1,\sigma}$ & 85/90/95 & 85.1/91.0/94.9 & 86.4/90.6/94.7  & 90.1/93.5/96.5 \\ 
   10 & 0.05 & $A^{i,1,2,\sigma}$ & 85/90/95 & 85.3/89.9/94.6 & 86.1/90.9/95.4  & 90.1/93.1/97.2  \\ 
   15 & 0.05 & $A^{i,1,2,\sigma}$ & 85/90/95 & 85.4/89.8/95.4 & 85.9/90.5/94.9  & 90.3/93.0/96.7  \\ 
   30 & 0.05 & $A^{i,1,2,\sigma}$ & 85/90/95 & 85.0/90.2/95.6 & 85.9/89.8/94.9  & 90.2/92.9/96.6  \\ 
  10 & 0.05 & $A^{i,3,1,\sigma}$ & 85/90/95 & 84.8/90.0/95.3 & 86.2/90.9/95.5  & 91.0/93.6/97.1 \\ 
  15 & 0.05 & $A^{i,3,1,\sigma}$ & 85/90/95 & 84.3/89.9/95.2 & 86.2/90.6/95.0  & 90.3/93.0/96.2 \\ 
  30 & 0.05 & $A^{i,3,1,\sigma}$  & 85/90/95 & 84.7/90.1/95.2 & 86.6/90.8/94.9  & 90.0/92.6/96.5\\ 
  10 & 0.05 & $A^{i,3,2,\sigma}$  & 85/90/95 & 86.0/90.6/95.0 & 85.4/90.3/95.5  & 90.3/93.3/96.9\\ 
  15 & 0.05 & $A^{i,3,2,\sigma}$  & 85/90/95 & 84.9/90.0/94.7 & 85.4/90.5/95.3  & 90.1/93.5/97.3\\ 
  30 & 0.05 & $A^{i,3,2,\sigma}$  & 85/90/95 & 85.1/89.7/95.3 & 85.9/90.7/94.9  & 89.9/92.9/96.5\\ 
  10 & 0.1 & $A^{i,1,1,\sigma}$  & 85/90/95 & 84.7/90.8/94.9 & 85.2/91.4/95.4 & 90.3/93.4/96.7\\ 
  15 & 0.1 & $A^{i,1,1,\sigma}$  & 85/90/95 & 84.9/89.8/95.1 & 86.1/90.4/95.1 & 89.5/91.6/96.6\\ 
  30 & 0.1 & $A^{i,1,1,\sigma}$  & 85/90/95 & 85.0/90.5/95.1 & 85.8/91.1/95.5 & 89.9/92.7/96.3\\ 
  10 & 0.1 & $A^{i,1,2,\sigma}$  & 85/90/95 & 85.5/90.4/94.5 & 86.3/90.8/95.1 & 90.3/93.3/96.4\\ 
  15 & 0.1 & $A^{i,1,2,\sigma}$  & 85/90/95 & 85.4/89.9/94.7 & 86.1/89.9/95.3 & 89.9/93.1/95.9\\ 
  30 & 0.1& $A^{i,1,2,\sigma}$  & 85/90/95 & 85.1/89.6/95.0 & 85.4/90.7/95.7 & 89.9/93.1/96.4\\ 
  10 & 0.1 & $A^{i,3,1,\sigma}$  & 85/90/95 & 85.4/90.1/96.0 & 85.4/90.2/94.6 & 90.1/93.6/97.0\\ 
  15 & 0.1 & $A^{i,3,1,\sigma}$  & 85/90/95 & 84.1/89.6/94.7 & 86.0/90.5/95.0 & 88.9/92.9/96.5\\ 
  30 & 0.1 & $A^{i,3,1,\sigma}$  & 85/90/95 & 85.4/90.3/94.9 & 85.3/90.1/95.3 & 88.9/93.4/96.5\\ 
  10 & 0.1 & $A^{i,3,2,\sigma}$  & 85/90/95 & 84.6/90.5/95.1 & 86.5/91.0/95.3 & 89.9/93.4/96.3\\ 
  15 & 0.1 & $A^{i,3,2,\sigma}$  & 85/90/95 & 85.2/90.2/95.1 & 86.2/89.8/95.3 & 89.8/93.1/96.2\\ 
  30 & 0.1 & $A^{i,3,2,\sigma}$  & 85/90/95 & 85.7/89.6/95.0 & 85.1/90.6/95.5 & 90.9/93.2/96.6\\ 
  10 & 0.6 & $A^{i,1,1,\sigma}$  & 85/90/95 & 82.4/87.7/93.9 & 81.6/87.3/93.6 & 87.1/91.2/95.5\\ 
  15 & 0.6 & $A^{i,1,1,\sigma}$  & 85/90/95 & 79.9/85.7/92.7 & 80.7/86.4/92.9 & 85.2/90.2/94.6\\ 
  30 & 0.6 & $A^{i,1,1,\sigma}$  & 85/90/95 & 79.4/85.5/92.4 & 78.7/84.8/92.3 & 82.8/87.6/92.9\\ 
  10 & 0.6 & $A^{i,1,2,\sigma}$  & 85/90/95 & 81.5/87.7/93.8 & 82.0/88.6/93.8 & 88.1/92.1/96.0\\ 
  15 & 0.6 & $A^{i,1,2,\sigma}$  & 85/90/95 & 81.9/86.8/93.1 & 81.0/87.1/93.2 & 86.3/90.5/94.7\\ 
  30 & 0.6 & $A^{i,1,2,\sigma}$  & 85/90/95 & 80.0/85.7/91.9 & 80.9/85.6/92.1 & 85.2/87.6/93.9\\ 
  10 & 0.6 & $A^{i,3,1,\sigma}$  & 85/90/95 & 83.0/88.7/94.7 & 84.2/88.8/94.2 & 88.1/91.6/96.0\\ 
  15 & 0.6 & $A^{i,3,1,\sigma}$   & 85/90/95 & 81.9/88.5/93.5 & 80.9/87.2/93.8 & 86.0/90.5/95.1\\ 
  30 & 0.6 & $A^{i,3,1,\sigma}$  & 85/90/95 & 80.2/86.7/93.1 & 80.0/86.3/92.8 & 85.0/89.5/94.0\\ 
  10 & 0.6& $A^{i,3,2,\sigma}$  & 85/90/95 & 84.3/89.7/94.4 & 84.2/89.0/94.9 & 87.4/92.5/96.2\\ 
  15 & 0.6 & $A^{i,3,2,\sigma}$ &  85/90/95 & 81.5/86.8/93.5 & 81.6/87.2/94.0 & 86.2/89.7/95.2\\ 
  30 & 0.6 & $A^{i,3,2,\sigma}$ &  85/90/95 & 81.3/86.6/92.4 & 81.8/86.7/92.4& 85.8/89.2/93.2\\ 
  \hline
\end{tabular}
\caption{Simulated covering rates (right box) of simultaneous $1-\alpha$-confidence tubes for GP models obtained from $M=5000$ simulations for varying error processes (E.P.). Notably, the Ornstein-Uhlenbeck processes ($i=3$) do not fulfill the assumptions necessary for application of the GKF.}
\label{tab:results}
\end{table}
  
  \FloatBarrier
  

\section{Application: Assessing Kneeling Effects on Gait}\label{scn:Application}

\paragraph{Study design.} 
In a study conducted at the School of Rehabilitation Science (McMaster University, Canada), 8 volunteers (4 female , 4 male,  for each gender, two aged 20-30 and two aged 50-60) with no previous knee injuries (external observation and subjective questioning revealed no obvious knee problems) with unremarkable knee kinematics motion have been selected.
In the experiment retro-reflective markers were placed onto identifiable skin locations on upper and lower volunteers' legs  by an experienced technician following a standard protocol. Eight cameras recorded the position of the markers and from their motions, a moving orthogonal frame $E_u(t)\in SO(3)$ describing the rotation of the upper leg w.r.t. the laboratory's fixed coordinate system was determined, and one for the lower leg, $E_l(t) \in SO(3)$, each of which was aligned near $I_{3\times 3}$ when the subject stood straight. 
As is common practice in clinical settings, subjects walked along a pre-defined 10 meter straight path at comfortable speed. For each of the following four sessions (A,B,C,D), for each subject a sample of $N\approx 12$ (for details on $N$, see Table \ref{N:tab})
repeated walks have been conducted and for every walk a single gait cycle $\gamma(t) = E_u(t) E_l(t)^T$ about half way through has been recorded, representing the motion of the upper leg w.r.t. the lower leg. After each walk the volunteers stopped shortly and started again for the next 10 meter walk. Thus, by design the assumption of independence of recorded gait cycles is satisfied.

\begin{table}[ht]
\centering
\fbox{\begin{tabular}{r|rrrrr}
 & 0\% & 25\% & 50\% & 75\% & 100\% \\
  \hline
A & 11.00 & 12.00 & 12.00 & 13.00 & 14.00 \\
  B & 12.00 & 12.00 & 13.00 & 13.25 & 14.00 \\
  C & 9.00 & 11.75 & 12.00 & 12.25 & 14.00 \\
  D & 9.00 & 11.00 & 12.00 & 12.25 & 13.00 \\
\end{tabular}}
\caption{\it \label{N:tab} Reporting the quartiles of numbers of processed walks (gait cycles) of volunteers for each of the four sessions from Table \ref{experiments:tab}.}
\end{table}

The study consists of four sessions, each giving, as described above, a sample of walks for the left leg of each volunteer. Between samples $A$ and  $B$ the markers were detached and placed again by the same technician following the same standard protocol. Hence the difference between these samples reflects the challenge of repeated reproducibility of gait patterns under clinical conditions. Before conducting the two sessions $C$ and $D$ markers where again replaced and the volunteers fulfilled a task of 15 minutes kneeling prior to data collection of session $C$ and yet another 15 minutes kneeling prior to session $D$. This allows to study the effect of kneeling and prolonged kneeling on gait patterns. Table \ref{experiments:tab} gives an overview of the four sessions conducted. Sessions A and B have already been reported in \cite{TelschowHuckemannPierrynowski2016}.

%
%
%

\begin{table}[h!]\centering
 \fbox{\begin{tabular}{c|l} Session& explanation\\ \hline
 A & no intervention, walks \\
 B & no intervention but marker replacement, walks\\
 C & marker replacement, 15 minutes of moderate kneeling, walks\\
 D & no marker replacement, another 15 minutes of prolonged kneeling, walks
 \end{tabular}}
\caption{\it \label{experiments:tab} Experiments conducted}
\end{table}

\paragraph{Dealing with the marker replacement effect.}
Replacing markers between sessions results in fixed and different rotations of the upper and lower leg, conveyed by suitable $P,Q \in SO(3)$ such that
$E^{\rm after}_u(t) = P E^{\rm before}_u(t)$ and $E^{\rm after}_l(t) = Q^T E^{\rm before}_l(t)$. Estimation of $P$ and $Q$ and temporal alignment of the sample mean curves have been done as described in Section \ref{scn:GPMs} and detailed in \citet{TelschowHuckemannPierrynowski2016} 
in order to make the samples comparable. Indeed, by Theorem \ref{theorem:EquivarianceConfidenceSets} the shape of the confidence sets does not depend on alignment correction.

In the following we report our findings, first in Table \ref{perm-test:tab} using the permutation test from \citet[Test 2.11]{TelschowHuckemannPierrynowski2016} correcting for sample-specific group action. If we were not to correct for sample-specific group action, we would detect significant changes of gait for 6 out of the 8 volunteers, even for ``A vs. B'', where nothing changed but marker placement, cf.  \citet[Table 2]{TelschowHuckemannPierrynowski2016} . The challenge dealt with in  \citet{TelschowHuckemannPierrynowski2016} was to design a test keeping the level, also under marker replacement.

\begin{table}[h!]\centering
\fbox{
 \begin{tabular}{c|cccc}
 Vol& A vs. C & B vs. C & A vs. D & B vs. D\\ \hline
 1 & 0.204& 0.158& \bf{0.029}&0.127\\
 2 & \bf{0.046}& \bf{0.002}& \bf{0.0}&\bf{0.0}\\
 3 & 0.872& 0.307& 0.191&0.311\\
 4 & \bf{0.001}& \bf{0.001}&\bf{0.0} &\bf{0.0}\\
 5 & 0.214& 0.735&0.559 &0.355\\
 6 & \bf{0.0}& \bf{0.0}& \bf{0.001}&\bf{0.008}\\
 7 & \bf{0.0}& \bf{0.0}&\bf{0.027} &\bf{0.042}\\
 8 & 0.467&0.705 &0.102 &0.149
 \end{tabular}} 

\caption{\it \label{perm-test:tab} Reporting $p$-values (significant in bold face) obtained from the permutation test in \citet[Test 2.11 in the version of Remark 2.12]{TelschowHuckemannPierrynowski2016} correcting for sample-specific group action.}

\end{table}

\paragraph{Results.} In Table \ref{perm-test:tab}, we see significant (often highly significant) changes of gait of volunteers $2$, $4$, $6$ and $7$ after each of the kneeling tasks. Volunteers $3$, $5$ and $8$ show no changes. Remarkably, these findings are consistent over marker replacement (``A vs. *'' and ``B vs. *'') and only for Volunteer $1$ the picture is unclear. 

\begin{figure}[h!]
  \centering
 \includegraphics[width=0.75\textwidth]{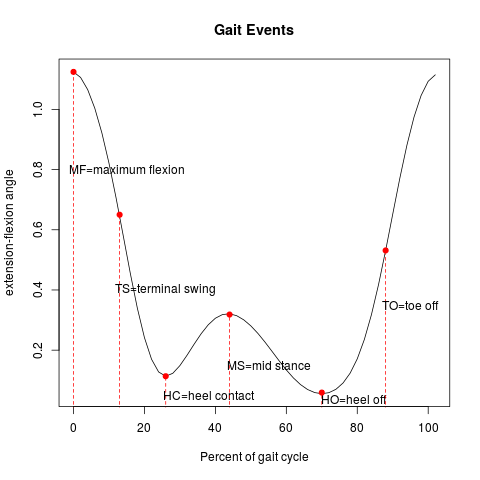}
\caption{\it \label{gait-event:fig} Depicting standard naming convention for gait events with respect to the flexion-extension angle.}
 \end{figure}

\begin{figure}[h!]
  \centering
 \includegraphics[width=1\textwidth]{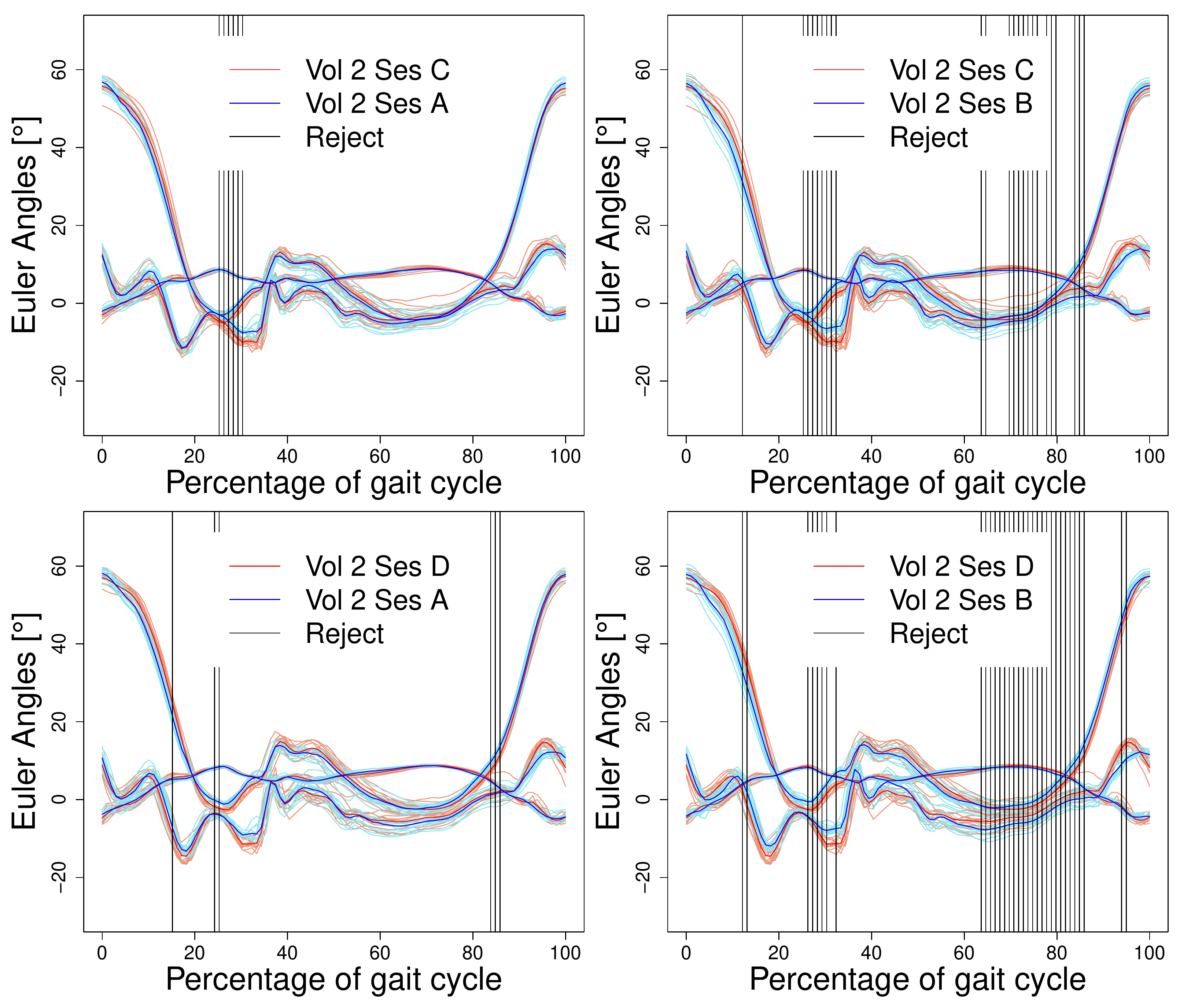}
 \caption{\it \label{conf-tubes-2:fig} Depicting for Volunteer 2 all three Euler angles of sampled gait curves for each of two different sessions. PEM curves are fat and vertical lines indicate loci of non-overlapping simultaneous $0.05$-confidence tubes in $SO(3)$. The largely varying curves are flexion-extension angles, cf. Figure \ref{gait-event:fig}, the middle curves with least variation are abduction-adduction and the bottom ones are internal-external angles.
}
 \end{figure}

\begin{figure}[h!]
  \centering
 \includegraphics[width=1\textwidth]{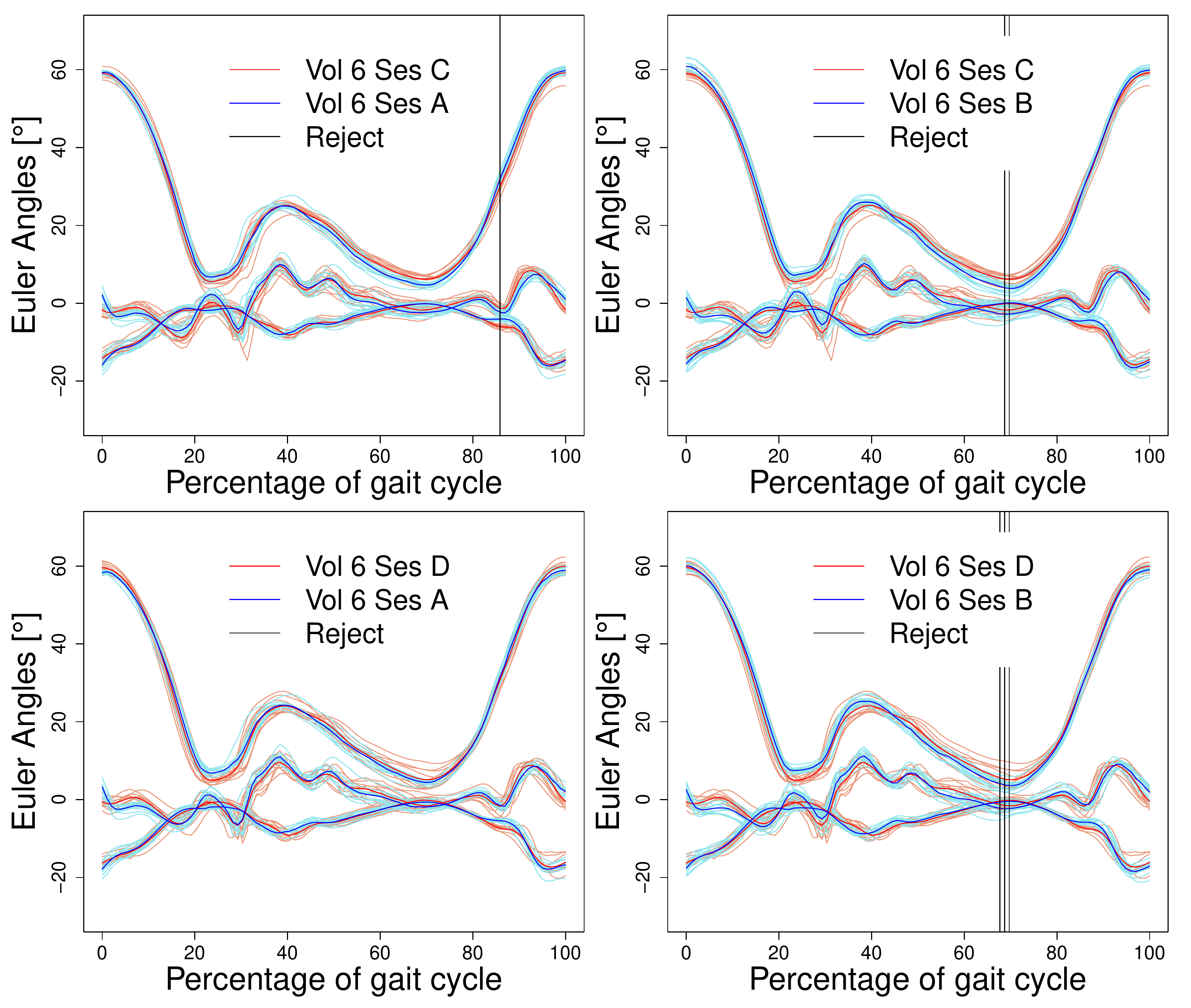}
 \caption{\it \label{conf-tubes-6:fig} Depicting with notation from Figure \ref{conf-tubes-2:fig} for Volunteer 6 all three Euler angles of sampled gait curves for each of two different sessions with PEM curves and loci of non-overlapping simultaneous $0.05$-confidence tubes in $SO(3)$.
}
 \end{figure}

In order to locate changes of gait patterns, we apply our new test of simultaneous confidence tubes. In Table \ref{kneeling:tab} we report the specific loci where $1-\alpha = 0.95$ confidence tubes no longer overlap, using standard naming convention (e.g. \cite{R95}) as illustrated in Figure \ref{gait-event:fig}. Employing Euler angles, which are popular in the field, as a local chart of $SO(3)$, the corresponding curves and specific loci of non-overlapping simultaneous confidence tubes are shown exemplary in Figures \ref{conf-tubes-2:fig} for Volunteer 2 and in Figure \ref{conf-tubes-6:fig} for Volunteer 6. Notably, non-overlapping confidence tubes have been determined in $SO(3)$ and not in chart coordinates so that the chart representations only serve as an approximate visualization of the real situation which we cannot visualize. The other volunteers' (1, 3, 4 and 7) curves with loci of non-overlapping confidence tubes are shown in the appendix in Figure \ref{conf-tubes-others:fig}. Again, we see that Volunteers 5 and 8 feature no changes in gait pattern. 
Volunteer 7 reported physical pain after post-kneeling walking. Indeed, high variation in gait patterns corresponding to session $D$ (red, in the left two displays of the bottom row in Figure \ref{conf-tubes-others:fig}) widened the corresponding confidence tubes such that changes of gait in Session D were not detected.

Combining Tables \ref{perm-test:tab} and \ref{kneeling:tab} and taking into account age and gender, we see that older age (volunteers with even numbers belong to age group 50 - 60) favors a kneeling effect over young age (volunteers with odd numbers belong to age group 20 - 30). As a surprise, the effect seems to be overall stronger for males. Having established a tool chain to study such effects, this experiment warrants larger studies.

 
\begin{table}[h!]\centering
\fbox{
 \begin{tabular}{c||c|c|c|c||c|c}
  Vol. &A vs. C & B vs. C &  A vs. D & B vs. D & gender & age group \\ \hline
  1   &    &  &  & MS & m & 20-30 \\
  2  &  HC  &  TS, HC, HO--TO& TS,HC,TO  & TS, HC, HO--TO, MF & m & 50-60\\
  3  &    &   & HC &  & f & 20-30\\
  4  &    &   & TS & TS& f & 50-60 \\
  5  &    &   &  &  & m & 20-30\\
  6  & TO   & HO  &   & HO & m & 50-60\\
  7   &    &  HC &  & & f & 20-30 \\
  8  &    &   &  &  & f & 50-60
 \end{tabular}}
 \caption{\it \label{kneeling:tab} Events from Figure \ref{gait-event:fig} where gait patterns changed such that $0.95$-confidence tubes no longer overlap.} 
\end{table} 

\section{Discussion}

In conjunction with the permutation test and estimation of marker replacement effects from \cite{TelschowHuckemannPierrynowski2016}, with the test for simultaneous non-overlapping confidence tubes presented in this paper, we have developed a tool chain that can be used in clinical practice to assess changes of gait patterns and localize these. 
These are no longer based on (single) Euler angle representations, as are often used in the field, but take advantage of a Gaussian perturbation model defined in the Lie group of three dimensional rotations. Due to the conservation of moment, gait curves are naturally smooth, their variation over repeated walks is moderate and hence approximations via the Gaussian kinematic formula are rather accurate, as well as in theory as in practice. 

In this study, with a small number of participants and a small number of repeated walks, we see that short kneeling tasks tend to affect gait patterns and it seems that older age and, possibly, male gender, favor this effect. We have made sure that this effect has not been caused by different marker placements. While specific loci of gait change depend on individuals, changes seem to occur least at local maxima of dominating flexion-extension, namely at MF and MS.

We believe that our results derived for $G=SO(3)$ generalize to general connected Lie groups setting as introduced in Sections \ref{scn:GPMs} and \ref{scn:Tubes}, in particular to products of $SO(3)$ with itself and with the Euclidean motion group, which are used in biomechanical analysis of more complicated joints (e.g. \citet{Rivest2008} for ankle motion) and in motion analysis of \emph{kinematic chains} of entire limbs (e.g. \citet{laitenberger2015refinement}) and their design for humanoid robots (e.g. \citet{ude2004programming}).
\FloatBarrier

\section{Acknowledgements}
 The first and the second author gratefully acknowledge support from DFG HU 1575/4 and 1575/7, the Niedersachsen Vorab
of the Volkswagen Foundation and DFG GRK 2088.

    \bibliographystyle{../../../../BIB/Chicago} 
\bibliography{../../../../BIB/shape,../../../../BIB/stochgeom,../../../../BIB/diffgeo,../../../../BIB/stats,../../../../BIB/biomech,../../../../BIB/numerics}

\newpage
\FloatBarrier
\appendix

\section{Appendix: More Visualizations of the Test for Non-Overlapping Confidence Tubes}\label{scn:Appendix}

 \begin{figure}[h!]
 \includegraphics[width=0.5\textwidth]{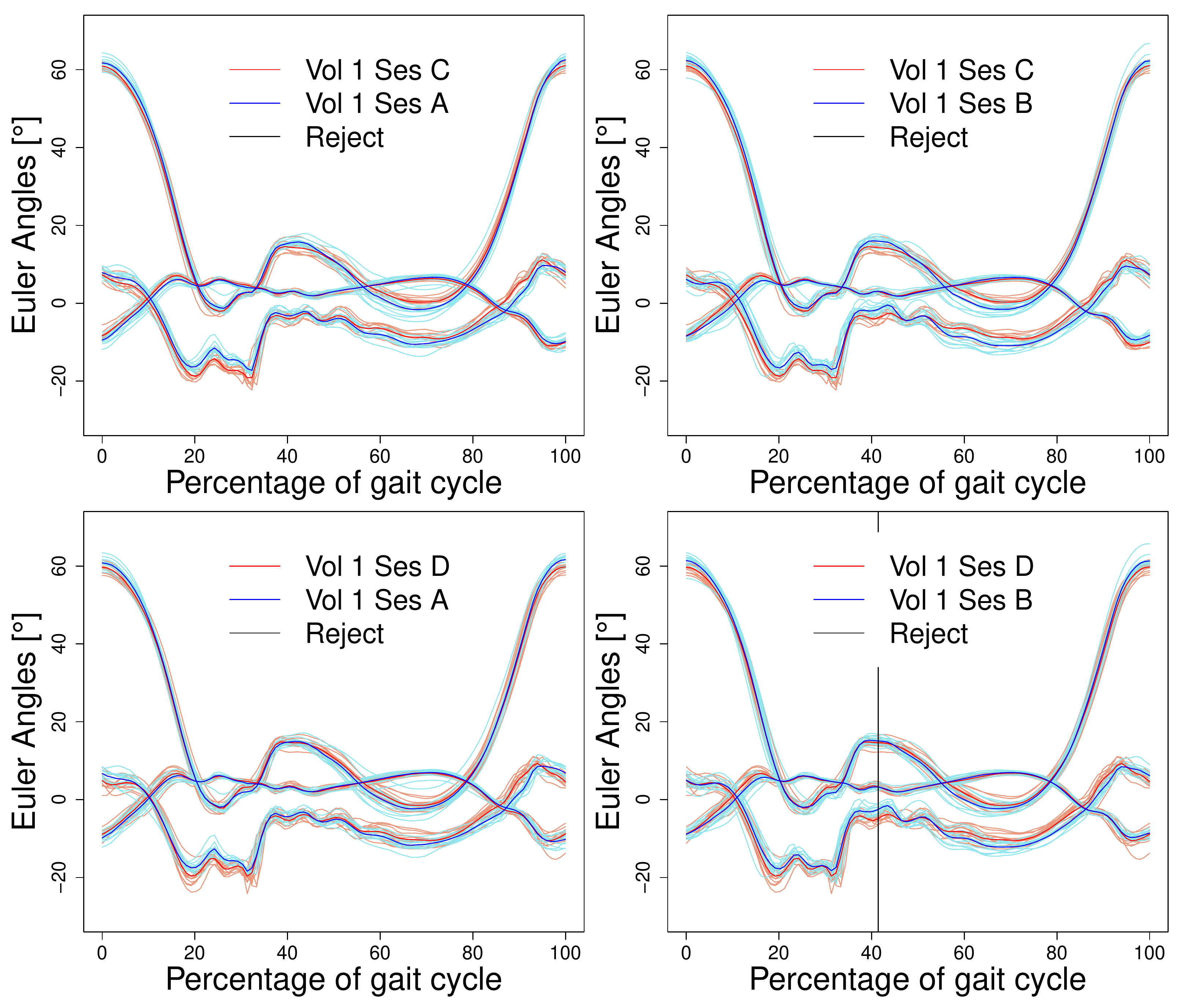}
 \includegraphics[width=0.5\textwidth]{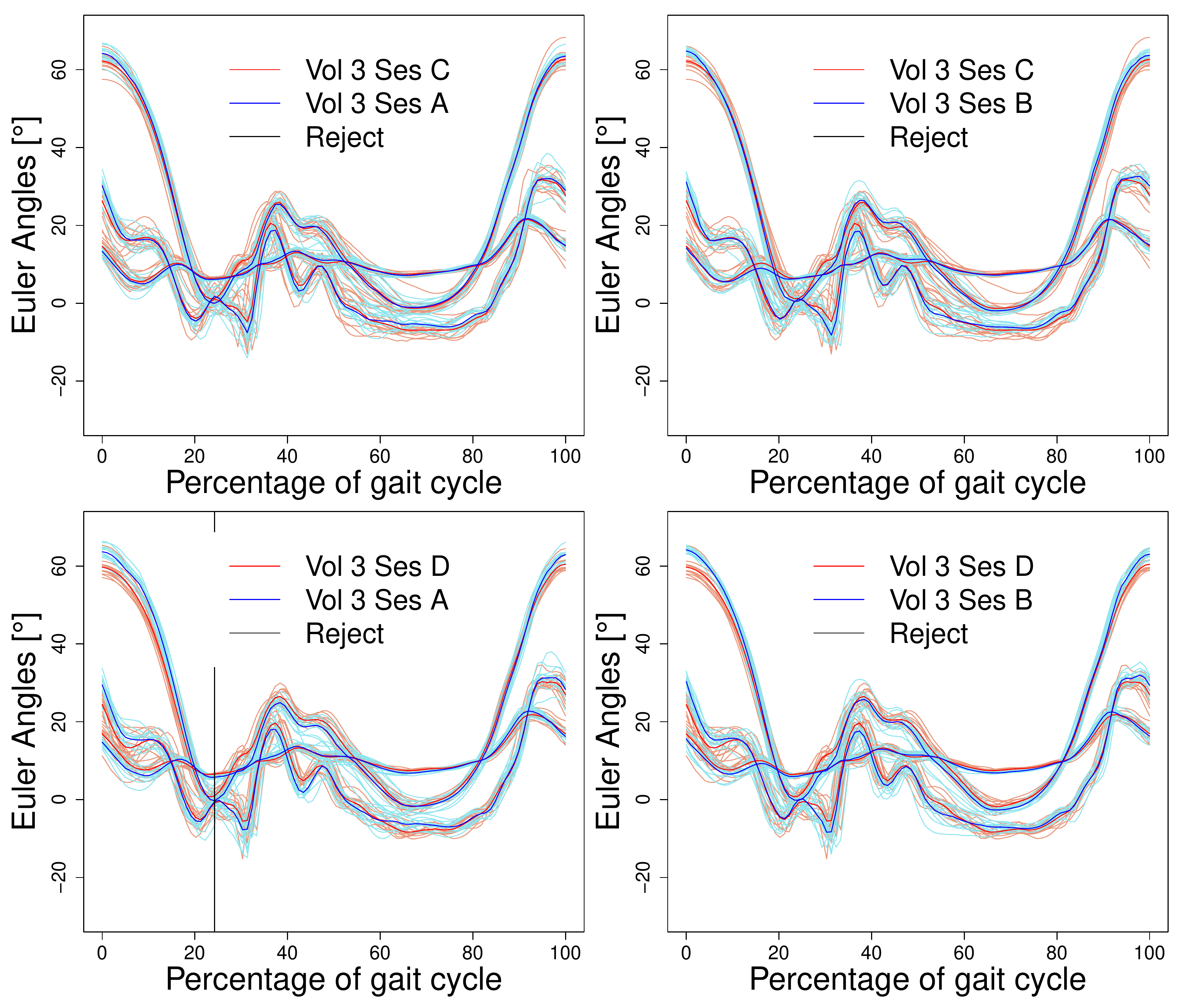}
 \includegraphics[width=0.5\textwidth]{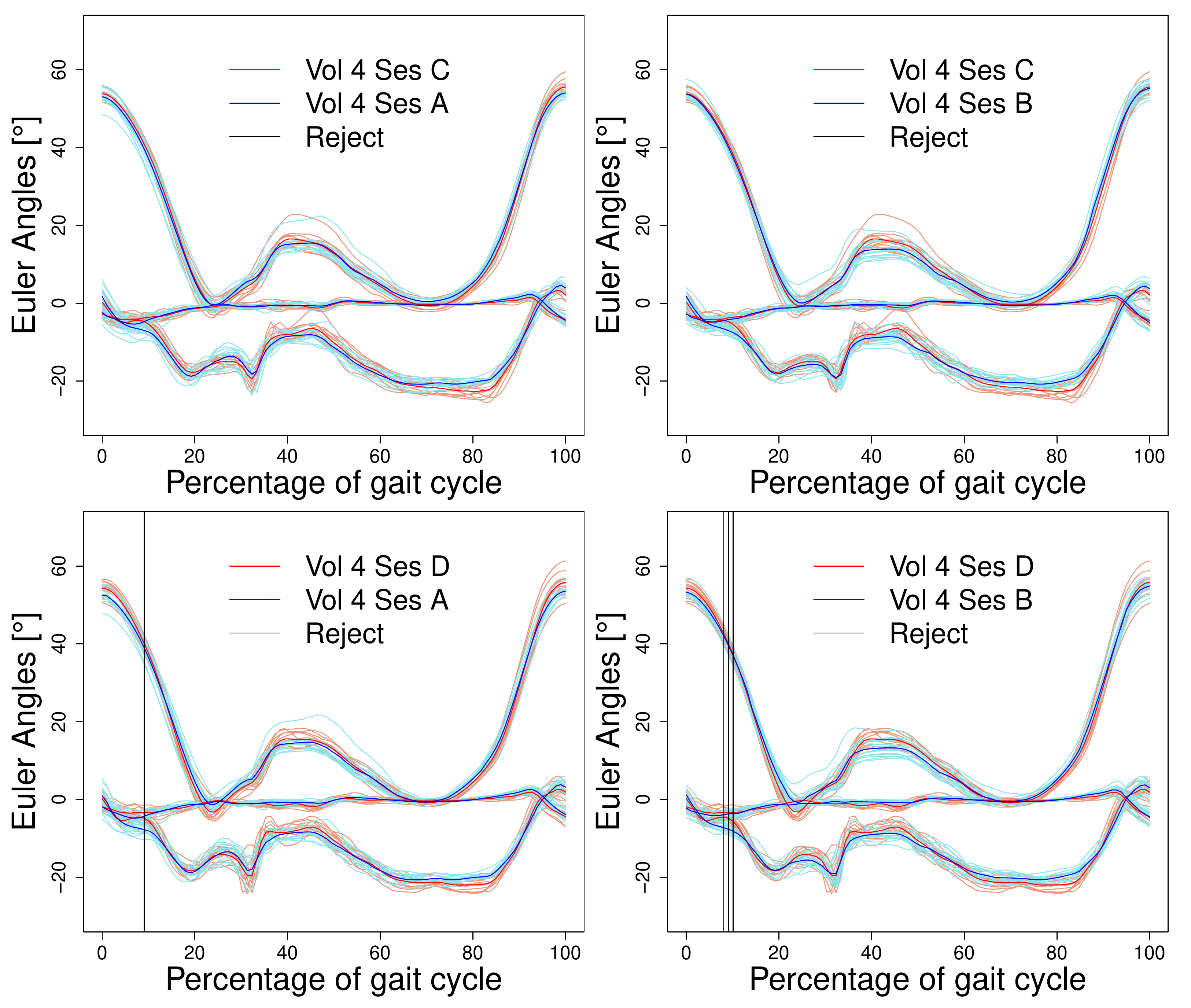}
 \includegraphics[width=0.5\textwidth]{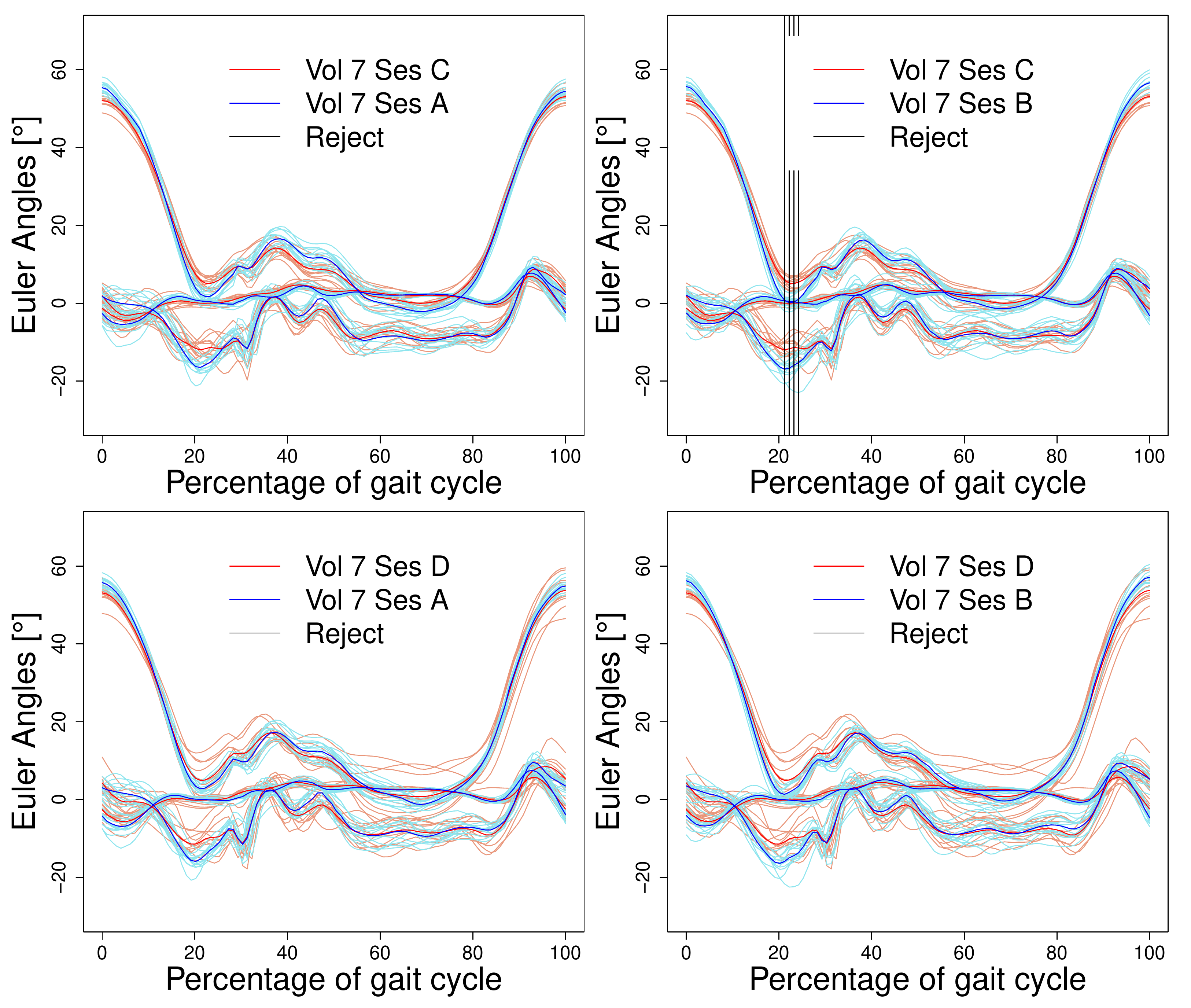}
 \caption{\it \label{conf-tubes-others:fig} Depicting with notation from Figure \ref{conf-tubes-2:fig} for Volunteers 1, 3, 4 and 7 all three Euler angles of sampled gait curves for each of two different sessions with PEM curves and loci of non-overlapping simultaneous $0.05$-confidence tubes in $SO(3)$.
}
 \end{figure}

\FloatBarrier

\section{Appendix: Proofs}\label{scn:Appendix}

\paragraph{Proof of Theorem \ref{theorem:residuals}}
Consider samples $\gamma_1,\ldots,\gamma_N$ with fixed $N\in \N$ of a GP model $\gamma_0 \Exp(A_t)$ with $a_t=\iota^{-1} \circ A_t$, $\max_{t\in [0,1]} \Vert a_t \Vert = \mathcal{O}_p(\sigma)$ and  $\sigma\rightarrow 0$, and let $\hat\gamma_{N}(t)\in E_N(t)$ be a measurable selection of PEMs.
Then for each $t\in [0,1]$, taking $x^N_t \in \R^3$ from Definition \ref{def:ResidualProcess} and $X^N_t = \iota \circ x_t^N$ we have that $\hat\gamma_{N}(t) = \gamma_0\,\Exp(X^N_t)$. Moreover, making use of the fact
\begin{equation}\label{eq:iota-product}
  \left( \iota\left(\frac{x}{\Vert x \Vert}\right) \right)^2 = \frac{xx^T}{\Vert x \Vert^2} - I_{3\times 3}\,.
\end{equation}
the property,
	$ \tr\big(\iota(c)^T\iota(d)\big) = 2c^Td$ for all $c,d\in \R^3$, and the Rodriguez formula (\ref{eq:rodriguez}), we have  for each $t\in [0,1]$ that $X^N_t$ maximizes
	\begin{align*}
	\frac{1}{N}\sum_{n=1}^N \tr&\Bigg( \hat\gamma_{N}^T(t) \gamma_0(t)\Exp\!\left(  A^{n}_t\right) \Bigg)	\\
	  =& ~ \tr\Bigg( \left(I_{3\times 3} +  \iota\big( x^N_t\big) \,\sinc\big(\big\Vert X^N_t\big\Vert_F\big) + \frac{1-\cos\big(\big\Vert X^N_t\big\Vert_F\big)}{\big\Vert X^N_t \big\Vert_F^2}\, \iota(x^N_t)^2\right)^T \\
	   &  \cdot \Big(I_{3\times 3} + \iota\big(\bar{a}^N_t\big) + \mathcal{O}_p\left( \sigma^2 \right)	\Big)\Bigg)\\
	   =&~ 3 + \tr\Bigg( \iota\left(x_t^N\right)^T~ \iota(\bar a^N_t) ~\sinc(\|x^M_t\|) + (1-\cos(\|x^N_t\|) ~\left(\iota\left(\frac{x^N_t}{\Vert x^N_t \Vert}\right)\right)^2 + \mathcal{O}_p(\sigma_l^2)\Bigg)\\
	   =&~ 1 + 2\left( {x_t^N}^T\bar a^N_t\,\sinc(\|x^N_t\|) + \cos(\|x^N_t\|)\right) + \mathcal{O}_p(\sigma^2)
	  \,.
	\end{align*}
	Note that the $\mathcal{O}_p\left( \sigma^2 \right)$ is indeed uniform in $t\in [0,1]$.
	
	Writing $x^N_t=re$ with a unit vector $e$ and length $0\leq r \leq \pi$, the first two summands above are maximized in $x^N_t$ if
	\begin{equation*}
	s\sin(r)  + \cos(r)
	\end{equation*}
	is maximal under the side condition $-\Vert \bar{a}^N_t \Vert \leq s = e^T\bar{a}^N_t \leq \Vert \bar{a}^N_t \Vert$. Hence, for $0\leq r<\pi$ choose the maximizing $s=\Vert \bar{a}^N_t \Vert$ (as large as possible) and hence $r=\arctan\!\big(\Vert\bar a^N_t\Vert\big)\in (0,\pi/2)$ ($r=\pi$ is no option). In consequence we have that
	\begin{equation*} 
	 x^N_t = \bar a^N_t \,\frac{\arctan\|\bar a^N_t\|}{\|\bar a^N_t\|} + \mathcal{O}_p\left(\sigma^2 \right)= \bar a^N_t + \mathcal{O}_p\left(\sigma^2 \right) \,.
	\end{equation*}
	This is \eqref{eq:residualsClaim1}.
	
	To establish equation \eqref{eq:residualsClaim2} from the above, consider the Taylor expansion
	\begin{align*}
	x^{N,n}_t= \mathfrak{L}\Big(\hat \gamma_{N}^T(t) \gamma_{n}(t)\Big) 	&= \iota^{-1}\circ\Log \Bigg( \Exp\Big(-\iota \circ \bar a^N_t  + \mathcal{O}_p\big(\sigma^2\big)\Big) \, \Exp\Big(\iota \circ  a^{n}_t\Big)\Bigg)\\
								  &= a^{n}_t -  \bar a^{N}_t + \mathcal{O}_p(\sigma^2) 
	\end{align*}
	wich is not valid for $\|a^n_t-x^N_t\| \geq \pi$, cf. \citet[p. 121]{ChirikjianKyatkin2000}. The probability of which, however, is $\mathcal{O}(\sigma^2)$, uniformly over $t\in [0,1]$, yielding the second assertion.
	
%
\qed

\paragraph{Proof of Corollary \ref{corollary:Tapproximation}}
Recall the definitions
\begin{equation*}
 H^{a,N}_t = N\big(\bar a^N_t\big)^T \!\left(S^{a,N}_t\right)^{-1} \!\bar a^N_t ~ ~ ~\text{ and } ~ ~ ~ \tilde H^{x,N}_t = N\big(x^N_t\big)^T \!\left(\hat S^{x,N}_t\right)^{-1} \!x^N_t\,.
\end{equation*}
By virtue of Theorem \ref{theorem:residuals} we obtain
\begin{equation*}
 \hat S^{x,N}_t = S^{a,N}_t + Z_t
\end{equation*}
with $\max_t \big\Vert Z_t \big\Vert_F = \mathcal{O}_p(\sigma_l^3)$. Using \citet[p. 58, eq. (24)]{HendersonSearle1981} yields
\begin{align*}
	\frac{1}{N} \hat H^{x,N}_t &= \big(x^N_t\big)^T \!\left(S^{a,N}_t + Z_t\right)^{-1} \!x^N_t\\
				&= \big(x^N_t\big)^T \!\left(S^{a,N}_t\right)^{-1} \!x^N_t - \big(x^N_t\big)^T \!\left(S^{a,N}_t\right)^{-1}\!Z_t\!\left( I_{3\times 3}+ \left(S^{a,N}_t\right)^{-1}\!Z_t \right)^{-1} \!\left(S^{a,N}_t\right)^{-1}\!x^N_t\,.\\
\end{align*}
From the assumption $\var\big[a^n_t\big]=\sigma^2\Sigma_t$ we have that $\max_{t\in [0,1]}\Big\Vert \big(S^{a,N}_t\big)^{-1} \Big\Vert_F = \mathcal{O}_p(\sigma^{-2})$.
Thus, we obtain
\begin{equation*}
    \big(x^N_t\big)^T \!\left(S^{a,N}_t\right)^{-1} \!x^N_t = \frac{1}{N} H^{a,N}_t + \mathcal{O}_p(\sigma)
\end{equation*}
by equation \eqref{eq:residualsClaim1}. Moreover, we obtain that $\max_{t\in [0,1]}\Big\Vert \left(S^{a,N}_t\right)^{-1}\!Z_t \Big\Vert_F =\mathcal{O}_p(\sigma)$ implying $\left(S^{a,N}_t\right)^{-1}\!Z_t \xrightarrow \Prb 0$ uniformly over $t\in[0,1]$. 
In consequence, on  $U=\left\{ \Big\Vert \left(S^{a,N}_t\right)^{-1}\!Z_t \Big\Vert_F < 1 \right\}$
we have the Von Neumann series
\begin{equation*}
      \left( I_{3\times 3} + \left(S^{a,N}_t\right)^{-1}\!Z_t \right)^{-1} = \sum_{j=0}^\infty (-1)^j \left( \left(S^{a,N}_t\right)^{-1}\!Z_t \right)^{j}
\end{equation*}
showing at once
\begin{equation*}
      \big(x^N_t\big)^T \!\left(S^{a,N}_t\right)^{-1}\!Z_t\!\left( I_{3\times 3}+ \left(S^{a,N}_t\right)^{-1}\!Z_t \right)^{-1} \!\left(S^{a,N}_t\right)^{-1}\!x^N_t =  \mathcal{O}_p(\sigma)\,.
\end{equation*}
Since $\Prb\{U\} = 1 - \mathcal{O}(\sigma)$, this completes the proof.
\qed

\paragraph{Proof of Theorem \ref{theorem:EquivarianceConfidenceSets}.}

      With the intrinisic residuals for each of the samples:
      \begin{equation*}
	  x^{N,n}_t = \mathfrak{L}\Big(\hat\gamma_{N}^T(t) \gamma_{n}(t)\Big) ~ ~ ~ \text{ and } ~ ~ ~ y^{N,n}_t = \mathfrak{L}\Big(\hat \eta_{N}^T(t) \eta_{n}(t)\Big)\,,
      \end{equation*}
      due to equivariance, $\hat \eta_N = \psi \circ \hat \gamma_N\circ \phi$, setting $\psi(R) = P_\psi R Q_\psi$ with $R,P_\psi,Q_\psi\in SO(3)$, we have
      \begin{align*}
      y^{N,n}_t			&= \iota^{-1}\!\circ\Log\Big(Q_\psi\hat\hat_{N}^T\big(\phi(t)\big) \gamma_{n}\big(\phi(t)\big) Q_\psi \Big) \\
				&= \pm  \iota^{-1}\left( Q^T_\psi\,\Log\Big(\hat\gamma_{N}^T\big(\phi(t)\big) \gamma_{n}\big(\phi(t)\big) \Big)Q_\psi\right)\,\\
				&= \pm Q_\psi\, \iota^{-1}\!\circ\Log\Big(\hat\gamma_{N}^T\big(\phi(t)\big) \gamma_{n}\big(\phi(t)\big) \Big)\\
				&= \pm Q_\psi\, x^{N,n}_{\phi(t)}\,.
      \end{align*}
      Here, the second equality is due to the power series expansion of the matrix logarithm and the observation that different extensions of the matrix logarithm to the cut locus of $I_{3\times 3}$ differ only by their sign; the third equality is due to (\ref{iota-rotation:eq}). 
      Moreover, by a similar argument for $x^N_t = \mathfrak{L}\Big(\hat\gamma_{N}^T(t) \gamma_0(t)\Big)$ and $ y^N_t = \mathfrak{L}\Big(\hat \eta_{N}^T(t) \eta_0(t)\Big)$ we obtain $y^N_t=\pm Q_\psi x^N_{\phi(t)}$, yielding
      \begin{equation*}
       \hat S^{y,N}_t =Q_\psi\hat S^{x,N}_{\phi(t)}Q_\psi^T\,,\quad \hat H^{y,N}_t = \hat H^{x,N}_{\phi(t)}\quad \text{ and } \quad  \hat h_{\gamma,N,\beta} =  \hat h_{\eta,N,\beta}\,.
      \end{equation*}
    This implies $ \mathcal{V}_{\beta}\big((\eta_1,\ldots,\eta_N); t\big) = Q_{\psi}\mathcal{V}_{\beta}\big(\gamma_1,\ldots,\gamma_N; \phi(t)\big)$, yielding the assertion.

\qed

\end{document}